\documentclass{elsart}

\usepackage{amssymb}    
\usepackage{bm}     
\usepackage{dcolumn}    
\usepackage{graphicx}
\usepackage{subfigure}

\newlength{\myw}

\begin{document}
\begin{frontmatter}

\title{Constricted Boron Nanotubes}
¨
\author[stuttgart,hgw]{Jens Kunstmann\corauthref{cor1}},
\ead{j.kunstmann@fkf.mpg.de}
\author[hgw]{Alexander Quandt}

\address[stuttgart]{Max--Planck--Institut f\"ur Festk\"orperforschung,
 Heisenbergstra\ss e 1, 70569 Stuttgart, Germany }
\address[hgw]{Institut f\"ur Physik, Domstra\ss e 10a,
Ernst--Moritz--Arndt--Universit\"at Greifswald, 17489 Greifswald, Germany}
\corauth[cor1]{corresponding author}


\begin{abstract}
The recent discovery of pure boron nanotubes raises questions
about their detailed atomic structure. Previous simulations
predicted tubular structures with smooth or puckered surfaces.
Here, we present some novel results based on \textit{ab initio}
simulations of bundled single--wall zigzag boron nanotubes (ropes).
Besides the known smooth and puckered modifications, we found new
forms that are radially constricted, and which seem to be
energetically superior to the known isomers. Furthermore, those
structures might be interpreted as intermediate states between
ideal tubular phases and the known bulk phases based on boron
icosahedra.
\end{abstract}


\end{frontmatter}


\section{Introduction}

Besides the well known carbon nanotubes (CNTs) \cite{iijima:91}
there are various other inorganic materials forming nanotubular
compounds \cite{tremel_1999_acheint,tenne_2001_cnt}. A material
that was originally predicted by theory
\cite{boustani:97,gindulyte_1998_ic} are nanotubes made of pure
boron. The stability and the mechanical properties of boron
nanotubes (BNTs) should be quite similar to C-- and BN--nanotubes
\cite{boustani:99}. But from an electronic point of view, BNTs
should always be metallic, independent of their structure
\cite{boustani:99}, in contrast to CNTs, which are either
semiconducting or metallic, depending on their radius and
chirality \cite{dresselhaus:SFCN}. Another difference between CNTs
and BNTs is the potential of the latter to form covalent
intertubular bonds \cite{quandt:01}, whereas the CNTs may only
bind to each other via van der Waals types of interactions
\cite{dresselhaus:SFCN}. Very recently Ciuparu \textit{et al.}
\cite{ciuparu_2004_jpcb} successfully synthesized BNTs and thus
confirmed the suggested existence of BNTs, after similar efforts
had already lead to the discovery of novel types of boron nanorods
by various other groups
\cite{cao:01,wu_2001_admat,otten:02,zhang_2002_chemcom}.

Among other nanotubular materials which have already been
synthesized in the past, we just mention some of the most
prominent materials, which are BN \cite{chopra_1995_sci} or
MoS$_2$ \cite{tenne_1992_nat} nanotubes. Beyond that, some recent
theoretical studies point out the possible existence of a large
family of metal--boron nanotubes \cite{quandt:01,zhang_2002_prl},
or the structurally related CaSi$_2$ nanotubes
\cite{gemming_2003_prb}, which are still waiting for their
experimental verification.

\begin{figure}
  \subfigure(a){\includegraphics[height=0.088\textheight]{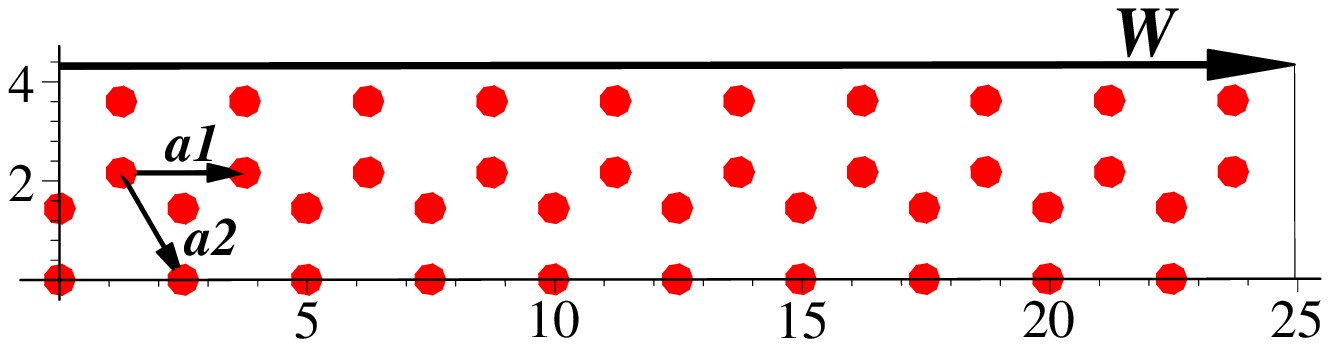}}
  \subfigure(b){\includegraphics[height=0.1\textheight]{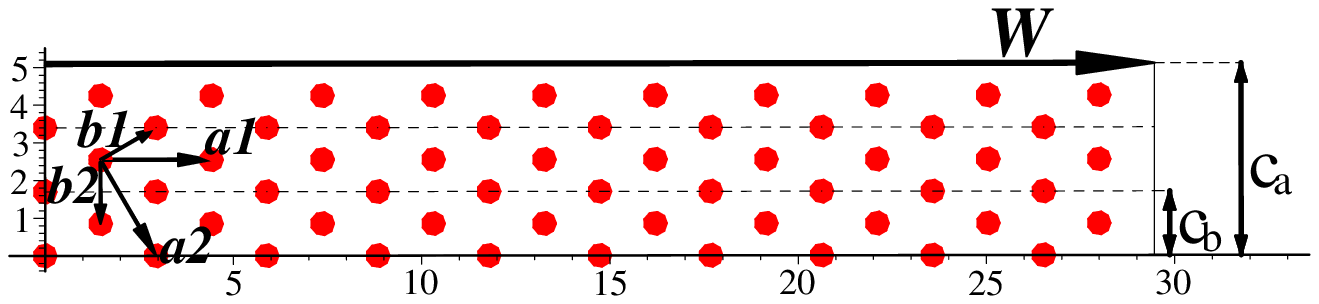}}
\caption{\label{fig:illu} Constructing the supercells for the
simulation of various nanotubes:  (10,0) sheets for (a) a carbon
nanotube and (b) a boron nanotube. Also shown are the wrapping
vector $\bm{W}$, the basis vectors $\bm{a_1}$ and $\bm{a_2}$ of a
honeycomb lattice and $\bm{b_1}$ and $\bm{b_2}$ of a hexagonal
lattice, and $c_{\mathrm a}$, $c_{\mathrm b}$, which are the
different heights of the supercells. The dashed lines indicate
that for zigzag systems $c_{\mathrm a} = 3 c_{\mathrm b}$. Units
are \AA.}
\end{figure}

Nanotubes are geometrically constructed by rolling up a
rectangular sheet that has been cut from a (quasi--)planar
structure. For CNTs this planar structure will be the honeycomb
lattice (Fig. \ref{fig:illu}a) \cite{dresselhaus:SFCN}, while for
BNTs the reference structure is a quasiplanar sheet, where the
boron atoms form a puckered hexagonal lattice (Fig.
\ref{fig:illu}b) \cite{boustani:99}. Due to differences in the
bond lengths ($a_{\mathrm{C-C}} = 1.44$ \AA \ and
 $a_{\mathrm{B-B}} \approx 1.65 \dots 1.85$ \AA) the boron and carbon sheets have
different sizes. All BNTs discussed in this paper are closely
related to the structure of CNTs, as they may be classified in a
standard fashion employing a pair of integers $(N,M)$. The latter
determine the so--called wrapping vector $\bm{W} = N \bm{a_1} + M
\bm{a_2} $, where $\bm{a_1}$ and $\bm{a_2}$ are basis vectors for
a \textit{honeycomb lattice}. It should be mentioned that BNTs may
as well be classified using the hexagonal basis vectors $\bm{b_1}$
and $\bm{b_2}$ (see Fig. \ref{fig:illu}b), but we certainly prefer
the use of the well--established classification scheme known from
CNTs. For a detailed discussion of the geometrical construction of
nanotubes see \cite{dresselhaus:SFCN}.

The existence of BNTs may be looked upon as special case of a more
general \textit{Aufbau principle} for boron clusters and bulk
materials proposed by Boustani \cite{boustani_1997_prb}. According
to this principle, stable {\em boron clusters} can be constructed
from two basic units: a pentagonal B$_6$ and a hexagonal B$_7$
pyramid. One structural paradigm is $\alpha$--boron, one of the
well--known bulk phases of pure boron, which is 'built' from
pentagonal pyramids forming B$_{12}$ icosahedra. In the bulk
phase, the icosahedral clusters occupy the vertices of a
rhombohedral unit cell, and the resulting structure is further
stabilized by complex multi--center bonds between the icosahedra.
As for the hexagonal pyramidal units, the \textit{Aufbau
principle} suggests that they may be combined to form convex or
quasiplanar clusters, which give rise to more complex
modifications, like spheres \cite{boustani_1997_jssc}, sheets
\cite{boust2:97}, or the above mentioned nanotubular forms of
boron.

In summary, the boron atoms in $\alpha$--boron have an inverse
umbrella six--fold coordination, while the tubular modifications
mainly exhibit a distinctive quasiplanar six--fold coordination. In
the following we will report about new modifications of zigzag boron
nanotubes, which might be interpreted as intermediate structures
between ideal nanotubular forms and the known bulk phases of pure
boron, emphasizing the general validity of the \textit{Aufbau
principle}.


\section{Theoretical Methods}

In order to simulate nanotube bundles (ropes), we have to
construct a solid composed of suitable supercells. Each supercell
contains a  single small boron ring. By piling up the supercells
in the $z$--direction (with lattice constant $c$) we build up an
infinite nanotubular structure. Within the $xy$--plane, we arrange
the nano\-tubes side by side on a hexagonal lattice with lattice
constant $a$.

The lattice constant $c$ depends on the chirality and the lattice
type of the nanotube \cite{dresselhaus:SFCN}, and it is
particular small for zigzag\footnote{Zigzag means $(N,M)=(N,0)$.}
systems. Figure \ref{fig:illu} illustrates that $c=c_a$ derived
from the honeycomb lattice (basis vectors $\bm{a_1}$ and
$\bm{a_2}$) leads to a supercell that is three times bigger than
one constructed from the basis vectors $\bm{b_1}$ and $\bm{b_2}$
($c_a = 3 c_b$). After a few simulations it turned out that the
results obtained from these two models are identical, and that all
known properties of BNTs can be well reproduced with the help of
the smaller supercells. Therefore most of the calculations where
performed with $c=c_b$ (see Tab. \ref{tab:data}).

Due to its electron deficient character \cite{pauling_1960_ncb}
boron has a complicated and versatile chemistry, as indicated
above. The only theoretical tools that allow to describe its
chemistry properly are first principles calculations
\cite{boustani_1997_prb}.

To this end, we used the VASP \textit{ab initio} package, version
4.4.5 \cite{kresse:96-1,kresse:96-2}. The latter is a density
functional theory \cite{kohn:65} based {\em ab initio} code
using plane wave basis sets and a supercell approach to model
solid materials. During all simulations, the electronic
correlations were treated within the local--density approximation
using the Perdew--Zunger--Ceperley--Alder exchange--correlation functional \cite{perdew_1981_prb,ceperley:80}, and the ionic cores of the system were
represented by ultrasoft pseudopotentials \cite{vanderbilt:90} as
supplied by G. Kresse and J. Hafner \cite{kresse_1994_jpcm}.

With the help of the VASP program, one can determine interatomic
forces and relax the different degrees of freedom for a given
decorated unit cell, and eventually detect atomic configurations
which correspond to (local) minima on the total energy
hypersurfaces. In order to carry out those structure optimizations
effectively, we employed a preconditioned conjugate gradient
algorithm \cite{teter:89} and allowed \textit{all} degrees of
freedom to relax (i.e. the complete set of atomic configurations
as well as the supercell parameters). The total energy and the
$k$--point sampling were converged such that changes in the total
energies were less than $10^{-3}$ eV and interatomic forces were
less than 0.04 eV/\AA.

The versatile chemistry of boron is reflected in a complicated energy
hypersurface, which is full of local minima. This is a particular
problem for structure
optimizations that aim at detecting structures corresponding to
global minima on those energy hypersurfaces, because the standard
techniques like the conjugate gradients method used in this study
are only able to find local minima.
In order to approach the most stable
structures, we developed a procedure where the starting
configurations were prerelaxed with lower numerical precision
until a (local) minimum was found, and afterwards we continued the
relaxations with optimal precision. A reduced precision leads to
somewhat imprecise interatomic forces, but we found that such a
procedure would result in the scanning of the energy hypersurface
over a wider range. This technique significantly
improves the optimizations, but the results still
depend on the starting configurations. Therefore we carefully examined each
system using different relaxation procedures \textit{and} a number
of distinct initial structures, with smooth or puckered surfaces.

We studied (6,0), (9,0), and (10,0) zigzag systems, and obtained
different isomers, which also vary in their cohesive
energies\footnote{The cohesive energy was calculated by dividing
the binding energy per supercell by the number of atoms contained
within that supercell.}. Depending on the structure of the surface
we discriminate between smooth (A), puckered (B) and constricted
(C) isomers. The related structural data, the parameters of the
supercells, the range of bond lengths and the cohesive energies of
all relaxed structures are given in table \ref{tab:data}. A top
view of all isomers placed around the center of their supercells
can be found in Fig. \ref{fig:(6,0)}--\ref{fig:(10,0)}. In these figures the big spheres stand for the
upper boron atoms and the small ones for the lower boron atoms
(with respect to the $z$--direction). The various lines between the
boron atoms point in the direction of the nearest neighbors
\footnote{Only nearest neighbor distances up to 2 \AA \ were
taken into account.}: a thin line symbolizes a single
neighbor, and a thick line two nearest neighbors.
According to the structures considered here and our experience
with other boron systems we found that it is chemically reasonable
to interprete distances $\le 1.9$ \AA \ as bonds of mainly covalent
character.

\begin{table}
\caption{\label{tab:data} Structural data and stabilities. $C_n$:
rotational symmetry, $n$: number of atoms per supercell,
$(a,b,c)_{h/m}$: lattice constants $a,b,c$ of the hexagonal or
monoclinic (index $h$ or $m$) supercell in \AA,
$a_{\mathrm{B-B}}^{\mathrm{intra}}$,
$a_{\mathrm{B-B}}^{\mathrm{inter}}$: range of bond lengths in \AA
\ of \textit{intra}tubular and \textit{inter}tubular bonds,
respectively, $E_{\mathrm{coh}}$: cohesive energies in eV/atom,
$\Delta E_{\mathrm{coh}}$: energies relative to the most stable
isomer of each system in eV/atom.}
 \newcolumntype{d}[0]{D{.}{.}{2.3}}
 \newcolumntype{e}[0]{D{,}{,}{5.5}}
 \begin{tabular}{cccccccdd}
 \hline \hline
 System  & Isomer & $C_n$ & $n$ & $(a,b,c)_{h/m}$ &
 $a_{\mathrm{B-B}}^{\mathrm{intra}}$ &
 $a_{\mathrm{B-B}}^{\mathrm{inter}}$ & \multicolumn{1}{c}{$E_{\mathrm{coh}}$} &
\multicolumn{1}{c}{$\Delta E_{\mathrm{coh}}$}\\
 \hline
 (6,0)  & A & $C_6$ & 12 & (8.18,8.18,1.65)$_h$ & $1.65 \dots 1.78$ & $-$ & 6.87 & - \\
 \hline
 (9,0)  & A & $C_3$ & 54 & (11.3,11.3,4.89)$_h$ & $1.63 \dots 1.80$ & $-$ & 6.81 & 0.19 \\
        & B & $C_3$ & 18 & (10.65,10.65,1.62)$_h$ & $1.62 \dots 1.81$ & $-$ & 6.95 & 0.05 \\
        & C & $C_3$ & 54 & (10.15,10.15,4.85)$_h$ & $1.62 \dots 1.86$ & 1.95 &7.00 & - \\
 \hline
 (10,0) & B & $C_2$ & 20 & (10.45,11.81,1.68)$_m$ & $1.68 \dots 1.90$ & $1.71, 1.96$ & 6.91 & 0.06 \\
        & C & $C_2$ & 20 & (11.46,10.46,1.64)$_m$ & $1.64 \dots 1.82$ & $1.84, 2.00$ & 6.97 & -\\
 \hline \hline
 \end{tabular}
\end{table}

The structure dubbed (6,0)A was prerelaxed with the Brillouin zone
being sampled by a 3$\times$3$\times$5 grid, and finished with a 5$\times$5$\times$11 grid.
Structure (9,0)A was completely relaxed on a 4$\times$4$\times$4 grid. 
Isomer (9,0)B was prerelaxed on a 2$\times$2$\times$9 mesh and completed on a 3$\times$3$\times$17 mesh.
For structure (9,0)C a 2$\times$2$\times$2 sampling was used for the
prerelaxation, and a 5$\times$5$\times$5 mesh to finish this simulation. 
A 5$\times$5$\times$11 grid was used to finish structure (10,0)B, with a 5$\times$5$\times$5
mesh used to carry out the prerelaxation. Finally (10,0)C was
prerelaxed with 3$\times$3$\times$5, and finished with a 5$\times$5$\times$11 grid. 
The cutoff
energy for the expansion of single electron wave function in terms
of plane waves was 257.1 eV for structures (9,0)B, and (10,0)C, and 321.4 eV
for all the other BNTs.


\section{Results}


A general feature of all nanotubes considered in this Letter is the
six--fold intratubular coordination with rather typical B--B bond
lengths (see Tab. \ref{tab:data}, column
$a_{\mathrm{B-B}}^{\mathrm{intra}}$). Furthermore all BNTs showed
a metallic density of states, confirming the results in
\cite{boustani:99}. The rotational $C_n$--symmetry of the tubes
follows from a simple rule: $n$ is the greatest common divisor of
the sixfold symmetry imposed through the arrangement on a hexagonal
superlattice and the $N$--fold symmetry of the BNT itself given by
the $(N,0)$ structure type.


\begin{figure}
\setlength{\myw}{0.6\textwidth}
\centering
\subfigure(A){\includegraphics[angle=-90,width=\myw]{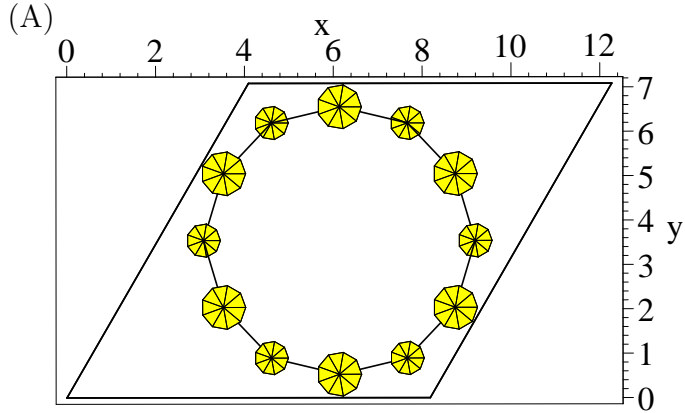}}

\caption{\label{fig:(6,0)}Top view of the only isomer of the (6,0)
zigzag system in its supercell. The big spheres stand for the
upper atoms and the small ones for the lower atoms (with respect
to the $z$--direction).}
\end{figure}

For the relaxation of the (6,0) system, we used different
relaxation schemes, and smooth or puckered initial structures, as
well as different supercell heights (see above). All structural
optimizations led to the same final structure, which is displayed
in Fig. \ref{fig:(6,0)}. It has a smooth surface and no
intertubular bonds. The radius of (6,0)A  is 3.03 \AA\ and the
intertubular distance is 2.12 \AA.


\begin{figure}
\setlength{\myw}{0.6\textwidth}
\centering
\subfigure(A){\includegraphics[angle=-90,width=\myw]{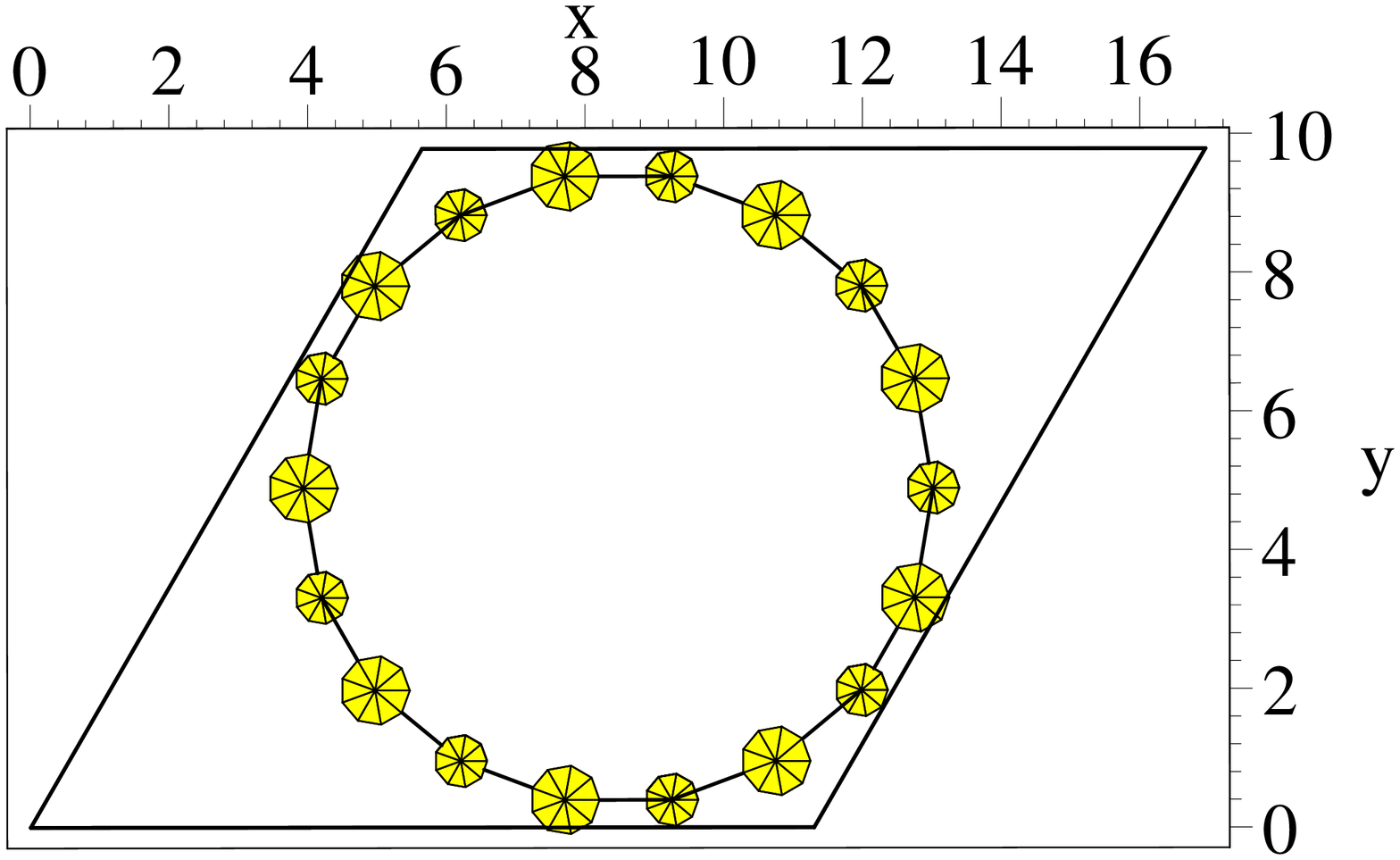}}

\subfigure(B){\includegraphics[angle=-90,width=\myw]{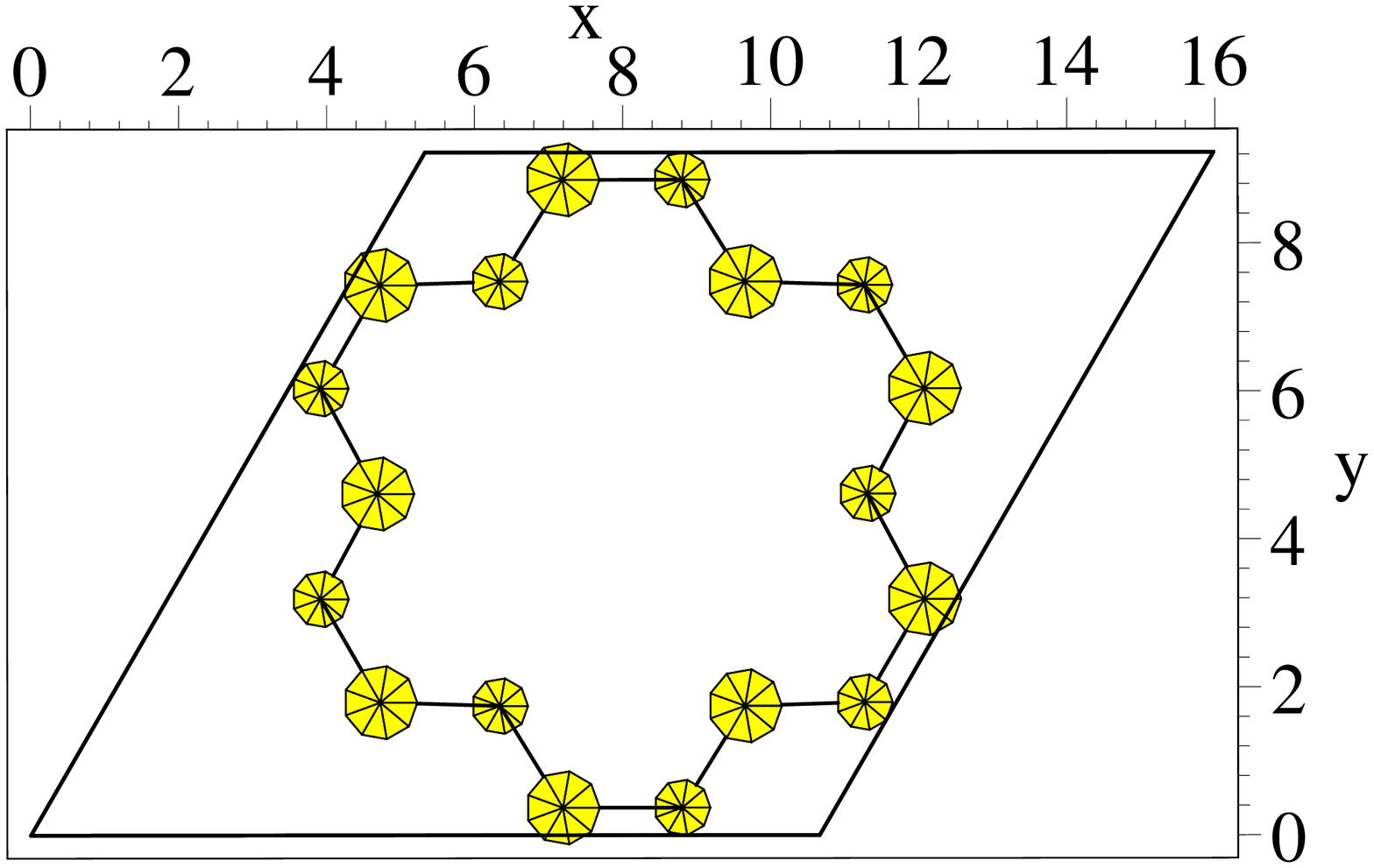}}

\subfigure(C){\includegraphics[angle=-90,width=\myw]{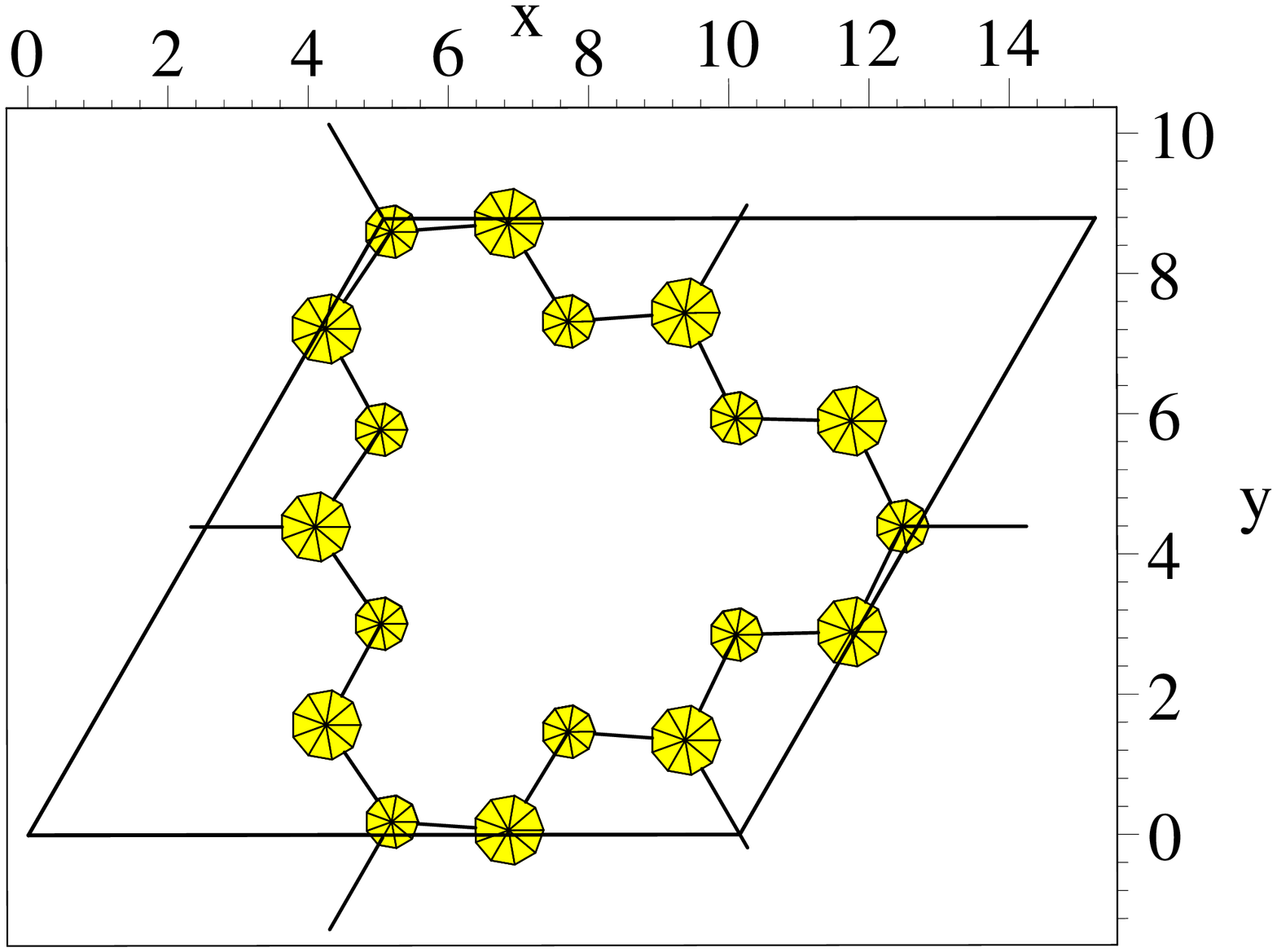}}

\caption{\label{fig:(9,0)}Top view of three isomers of the (9,0)
system placed around the center of their supercells. In (C), each
of the six additional thick lines point towards \textit{two} atoms
in the adjacent supercells within a distance of 1.95 \AA.}
\end{figure}

The energy surface of the (9,0) system seems to be more complex,
and we found three distinct isomers, which are displayed in Fig.
\ref{fig:(9,0)}. Similar to (6,0)A, structure (9,0)A has a smooth
surface and no intertubular bonds. The radius is 4.57 \AA\ and the
intertubular distance is 2.16 \AA. The second isomer (9,0)B has a
puckered structure described in \cite{boustani:99}, which results
from rolling up a regular quasiplanar surface on a cylinder. There
are no intertubular bonds; the closest distance of apex atoms to
an adjacent BNT is 2.63 \AA. The new modification (9,0)C is
qualitatively different from the previous ones, because it is
radially constricted, and cannot be generated by simply rolling up
a quasiplanar reference structure. The constriction generates
three bumps, each of them formed by five atoms. Furthermore there
are six atoms per supercell that have two neighbors in adjacent
supercells, which are 1.95 \AA\ apart (indicated by thick lines
pointing outwards in Fig. \ref{fig:(9,0)}C).


\begin{figure}
\setlength{\myw}{0.6\textwidth}
\centering
\subfigure(B){\includegraphics[angle=-90,width=\myw]{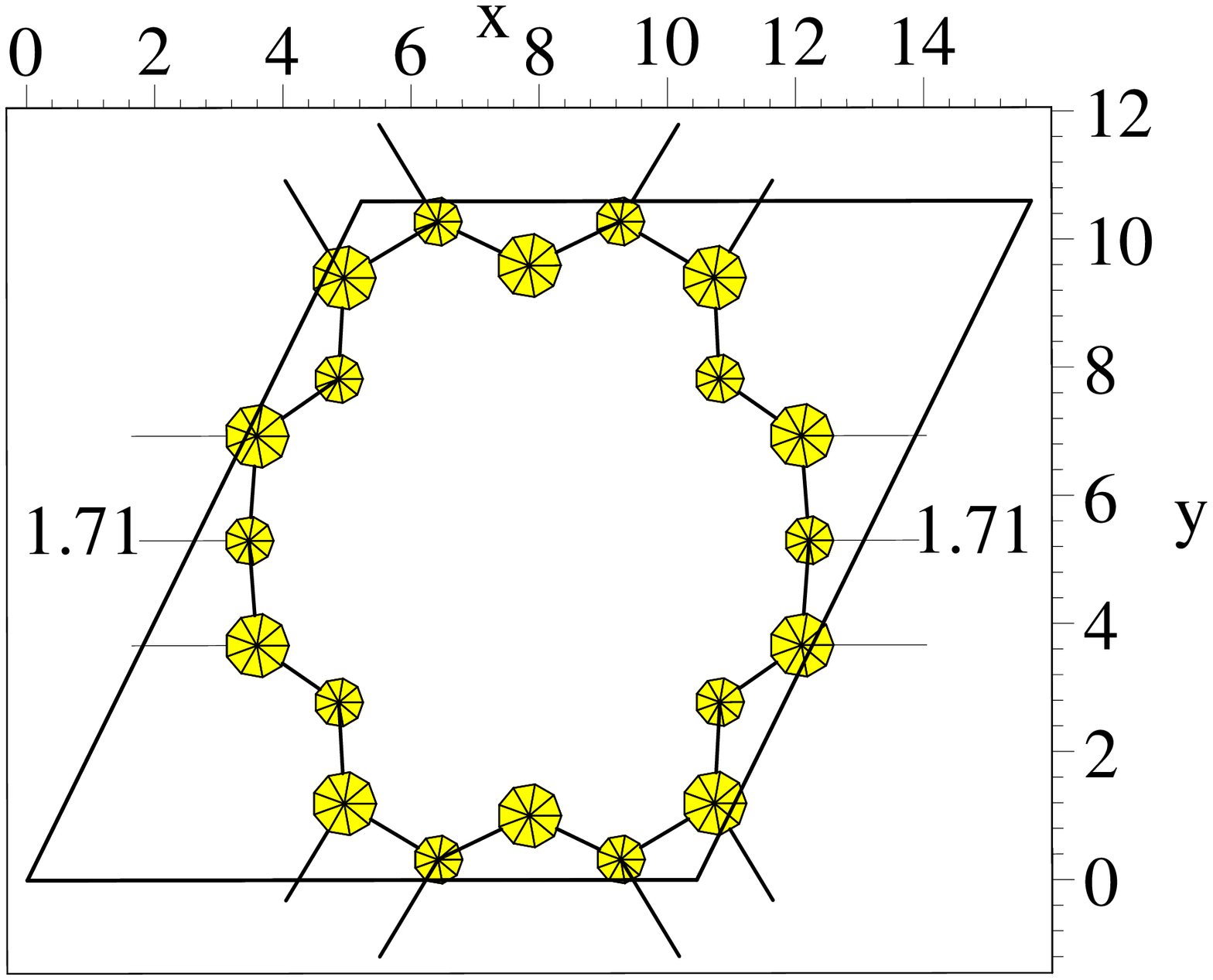}}

\subfigure(C){\includegraphics[angle=-90,width=\myw]{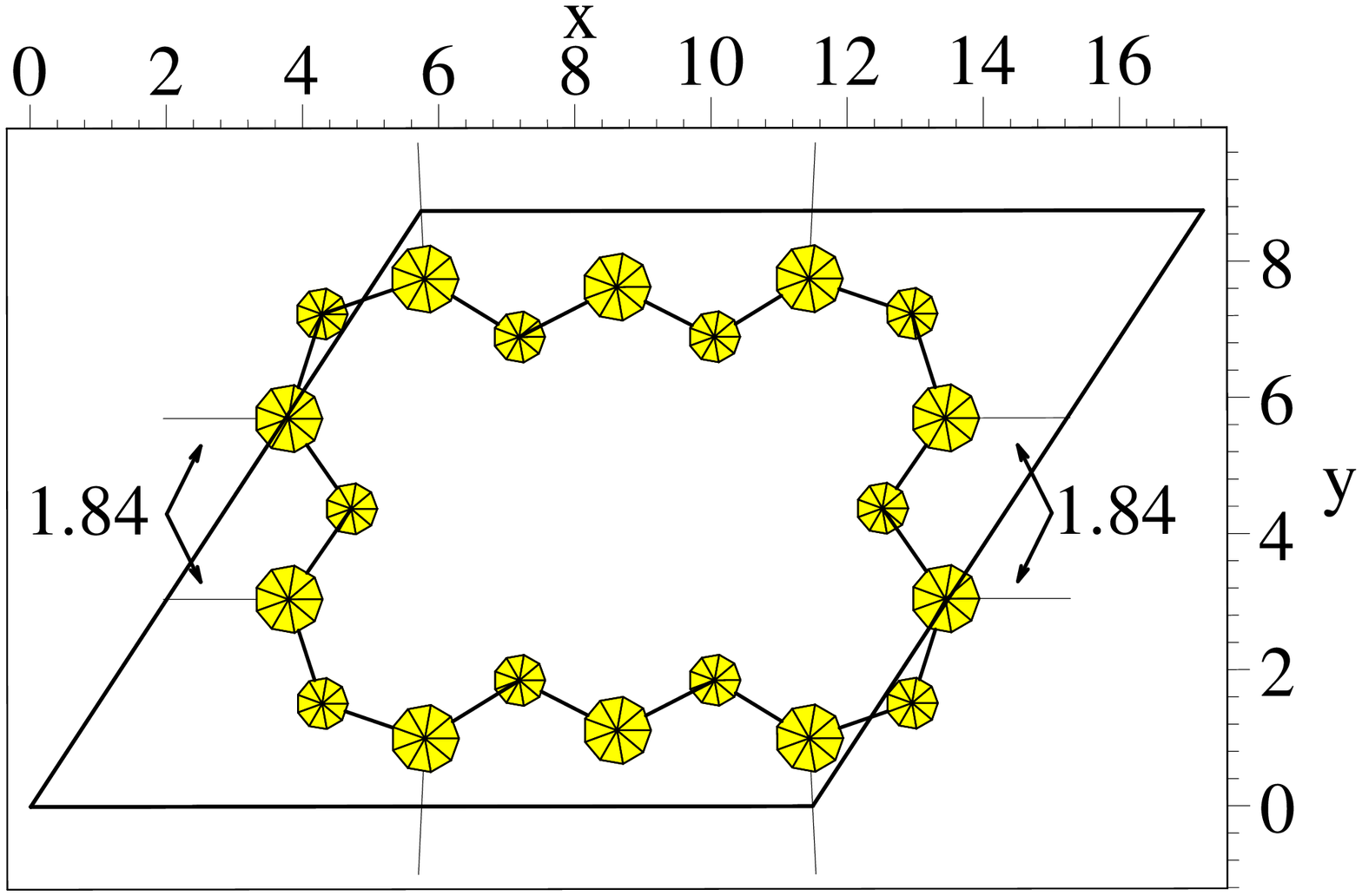}}

\caption{\label{fig:(10,0)} Two isomers of the (10,0) system in
top view located around the center of their their monoclinic
supercells. The additional lines point in the direction of nearest
neighbors: a thin line symbolizes a single neighbor, a thick line
two nearest neighbors. Isomer (B) has two intertubular bonds with
$a_{\mathrm{B-B}}=1.71$ \AA, whereas the remaining outward
pointing lines indicate atoms within a distance of 1.96 \AA.
Isomer (C) has four intertubular bonds ($a_{\mathrm{B-B}}=1.84$
\AA) and there are four atoms in neighboring supercells, which are
2.00 \AA\ apart.}
\end{figure}

For the (10,0) system we only found two isomers. It was not possible
to locate an A--type structure.
During the relaxation procedure the hexagonal symmetry of the
supercells was broken, and the lattice systems changed to
monoclinic.
The smaller angle of the (10,0)B supercell is 63.8$^{\circ}$, and
the one for (10,0)C is 56.7$^{\circ}$ (for a hexagonal supercell
this angle would be 60$^{\circ}$). Structure (10,0)B is puckered,
but its cross--section is elliptical and not circular as for the
other BNTs considered so far; it has two intertubular bonds per
supercell, which raise the coordination number of those atoms to
seven. Furthermore each BNT is surrounded by 20 atoms per
supercell, which belong to adjacent nanotubes that are 1.96 \AA\
apart (see the outward pointing lines in Fig. \ref{fig:(10,0)}B).
The cross--section of the constricted isomer (10,0)C  deviates
significantly from circular or elliptical shapes. The structure
has four bumps, each formed by five atoms. Furthermore it has four
intertubular bonds per supercell and four nearest neighbors in
adjacent supercells in a distance of 2.00 \AA.

The bumps found within the (9,0)C and (10,0)C structures are
formed by five boron atoms per supercell, which are sitting on the
corners of an imaginary zigzag 6--ring, similar to the six apex
atoms of a B$_{12}$ icosahedra seen along each of its 3--fold
axes. Therefore, the constricted nanotubes might be interpreted as
intermediate structures between the ideal (puckered) nanotubular
structures reported before and the well--known bulk phases of boron
based on B$_{12}$ icosahedra. The more as previous ab initio
studies \cite{boustani:96} showed that in principle, B$_{12}$
icosahedra may be fused to form extended nanotubular 6--ring
systems similar to the columnar wings of the (9,0)C and (10,0)C
structures.


\section{Conclusions}
We explored the geometry, energetics and basic chemical properties
of boron nanotube bundles (ropes) of zigzag type.

Our results confirm that zigzag BNTs tend to have puckered
surfaces \cite{boustani:99}. The (9,0)B isomer is clearly more
stable than the smooth A--type isomer. For the (6,0) system we
were unable to find a puckered isomer, while for the (10,0) system
we were unable to find a smooth BNT. From this one could certainly
conjecture that for small radii the surface tension is smoothening
the BNT, while for bigger radii there will be a strong tendency to
puckered modifications.
In that context it might be surprising that the (9,0)A isomer is
somewhat lower in binding energy than (6,0)A, which seems, at
first sight, to be in conflict with the strain energy studies in
\cite{boustani:97,boustani:99} (for a general discussion about
strain energy see \cite{dresselhaus:SFCN}).
But these studies generally referred to isolated BNTs, whereas in
the present Letter, we are dealing with ropes made from BNTs, which
show some complex intertubular bonding that is very likely to
alter the cohesive energies in a less predictable fashion.
Furthermore, by applying numerical optimization methods to complex
materials, one can never guarantee that the algorithm will really
succeed in locating the proper local minimum corresponding to a
smooth (9,0)A isomer.


Apart from that, the A-- and B--type isomers basically confirm
previous results known for (nanotubular) cluster systems
\cite{boustani:97,boustani:98,boustani:99}.
For periodic systems containing (9,0) and (10,0) zigzag boron
nanotubes, we found new types of radially constricted BNTs (C--type
isomers).
These isomers seem to be energetically favored over the previously known
BNT modifications (see column $\Delta E_{\mathrm{coh}}$ in Tab.
\ref{tab:data}), and their structures are qualitatively different
from all BNTs reported so far, because it cannot be derived from
simply rolling up a quasiplanar reference sheet on a cylinder.
The $C_n$ symmetry of these structures looks like a compromise
between the symmetry of the zigzag BNTs and the symmetry
constraints imposed by their arrangement within a hexagonal
tubular network (see above). Therefore we conjecture that the
constriction of the C--type isomers are most likely due to the
special boundary conditions imposed during the formation of
nanoropes, which lead to novel structures, quite different from
structures predicted for isolated BNTs.

Furthermore, the constriction of zigzag BNTs produces bumps formed
by five atoms per supercell sitting at the corners of a imaginary
6--ring, similar to structural motives found within B$_{12}$
icosahedra. Therefore, these structures may be looked upon as
intermediate structures between ideal nanotubular boron systems
and the known bulk phases of boron based on B$_{12}$--icosahedra,
confirming a general \textit{Aufbau principle} for boron
structures suggested by Boustani \cite{boustani_1997_prb}.

We hope that our results will inspire further experimental work on
BNTs going beyond the pioneering work of Ciuparu \textit{et al.}
\cite{ciuparu_2004_jpcb}. In particular, a metallic nanotubular
materials with versatile structural and chemical properties may
allow for many interesting applications within nanotechnology.

\section{Acknowledgments}
The authors thank S. Kosse (Greifswald) for technical support
during our extensive use of the 'snowwhite' computer cluster, K.
Fesser (Greifswald) for various helpful discussions and O. K.
Andersen (MPI Stuttgart) for supporting this work.

\bibliography{others,boron,nanowires,non-carbon,bcn,sulfides,numerics,ihsan+alex}

\end{document}